\documentstyle[aps,epsfig,twocolumn]{revtex}

\begin{document}

\draft 
\title{Intermediate-mass dilepton production in heavy-ion collisions
at 200A GeV}

\author{G.Q. Li and C. Gale\footnote{Permanent address: Physics 
Department, McGill University, Montreal, QC, H3A 2T8, Canada}}
\address{Department of Physics and Astronomy, State University of New
York at Stony Brook, Stony Brook, New York 11794}
\maketitle
  
\begin{abstract}
Through the analysis of HELIOS-3 data 
obtained at the CERN SPS, 
we demonstrate the importance of secondary processes 
for dilepton production in heavy-ion collisions in the
intermediate invariant mass region. We find that while the dilepton 
spectra between 1 to 2.5 GeV from proton-induced reactions 
can be attributed to the decay of primary vector mesons, 
charmed hadrons, and initial Drell-Yan processes; the strong 
enhancement seen in the heavy-ion data as compared to the
background comes mainly from the secondary processes
which are germane to heavy-ion collisions. Furthermore,
we find $\pi a_1\rightarrow l{\bar l}$ to be the most important 
process in this mass region, as was found by thermal rate
calculations. We emphasize the constraints on the elementary
cross sections by the experimental data from $e^+e^-$ annihilation.
\end{abstract}


The experimental measurement and theoretical investigation of
dilepton production in nuclear collisions constitutes one
of the most active and exciting fields in heavy-ion physics \cite{qm96}.
Because of their relatively weak final-state interactions
with the hadronic environment, dileptons are
considered ideal probes of the early stage of heavy-ion
collisions, where quark-gluon-plasma (QGP) formation and
chiral symmetry restoration are
expected \cite{shur80,brown96}.  

Dilepton mass spectra in heavy ion collisions can be 
divided into three regions.
The region below $m_\phi$ ($\sim$ 1 GeV)
is dominated by hadronic interactions and hadronic
decays at freeze-out. In the intermediate-mass region, 
$m_\phi < M < m_{J/\Psi}$, the contribution
from the thermalized QGP might be seen \cite{shur78}. 
In the high-mass region at
and above $m_{J/\Psi}$ the major effort has been the
detection and understanding of $J/\Psi$ suppression.
So far, the experimental measurement of dilepton spectra 
at the CERN SPS has mainly been carried out
by three collaborations: the CERES
collaboration has specialized in dielectron spectra in the 
low-mass region \cite{ceres,drees96}, the HELIOS-3 \cite{helios} 
collaboration has measured dimuon spectra from 
threshold up to the $J/\Psi$ region, and the NA38/NA50 \cite{na38} 
collaboration measures dimuon spectra in the intermediate- 
and high-mass regions.
 
Recent observation of the enhancement of low-mass dileptons in
central heavy-ion collisions over the proton-induced 
reactions by the CERES \cite{ceres,drees96} and
the HELIOS-3 \cite{helios} collaborations has generated a 
great deal of theoretical activity. The results from many 
groups with standard scenarios (i.e., using vacuum meson properties) 
are in remarkable agreement with each other, but in significant 
disagreement with the data: the experimental spectra in the 
mass region from 0.3-0.6 GeV are substantially underestimated 
\cite{drees96}. This has led to the suggestion of 
various medium effects that might be responsible for the
observed enhancement \cite{likob,rapp}. 

In the high-mass region around $m_{J/\Psi}$, the $J/\Psi$ suppression 
has been a subject of great interest, since it was first proposed
as a signal of the deconfinement phase transition \cite{satz86}. Various
investigations show that up to central S+U collisions, the normal
pre-resonance absorption in nuclear matter is sufficient to account 
for the observed $J/\Psi$ suppression. However,
recent data from the NA50 collaboration for central Pb+Pb
collisions show an additional strong `anomalous' suppression
which might indicate the onset of the color deconfinement
\cite{na50}. 

Another piece of interesting experimental data that has not received
much theoretical attention is the dilepton spectrum in the
intermediate-mass region between 1 and 2.5 GeV.
Both the HELIOS-3 and NA38/NA50 collaborations have observed
a significant enhancement of the dilepton yield in this mass region
in central S-induced collisions, as compared to that in the 
proton-induced reactions \cite{helios,na38}. Preliminary data from the
NA50 collaboration also show significant enhancement in central
Pb+Pb collisions \cite{na38} (see also Ref. \cite{drees96}).
 
The intermediate-mass dilepton spectra in heavy-ion collisions
are particularly useful for the search of the QGP \cite{shur78}. 
However, to extract from the 
measured dilepton spectra any information about the phase 
transition and the properties of the QGP, it is essential 
that the contributions from the hadronic phase be precisely 
understood and carefully subtracted.
It is the purpose of this work to calculate the dimuon spectra
in central S+W collisions based on the relativistic transport 
model used in Ref. \cite{likob,lib97} for low-mass dilepton and
photon production. The chief motivation for such a study is to
understand the origin of the observed enhancement in the 
intermediate-mass region, and to see whether the enhancement
can be explained by hadronic processes, or whether the formation of 
the QGP needs to be invoked.

Previous thermal rate calculations based on kinetic theory
show that in the mass and temperature region relevant for
this study, the following thermal processes (from the hadronic
phase) are important: $\pi\pi\rightarrow l{\bar l}$,
$\pi\rho\rightarrow l{\bar l}$, $\pi\omega\rightarrow l{\bar l}$,
$\pi a_1\rightarrow l{\bar l}$, $K{\bar K}\rightarrow l{\bar l}$, 
and $K{\bar K^*}+c.c \rightarrow l{\bar l}$ 
\cite{song94,haglin95,kim96}. 
Among them, the $\pi a_1\rightarrow l{\bar l}$ has 
been found to be the most important, mainly because of its
large cross section \cite{song94,kim96} (A similar conclusion
has been drawn for thermal photon production \cite{xiong92,song93}).
In this work we shall verify quantitatively the previous
estimates.

To compare with experimental data, 
one needs a transport model that describes
the dynamical evolution of the colliding system, and integrates 
the dilepton production over the entire reaction volume and time.
In heavy-ion collisions at CERN SPS energies, 
many hadrons are produced in the initial nucleon-nucleon 
interactions. This is usually modeled by the fragmentation of 
strings. One successful model 
for taking into account this non-equilibrium dynamics is the RQMD model 
\cite{sorge89}. As in Refs. \cite{likob,lib97}, we use as
initial conditions the hadron abundance and distributions 
obtained from the string fragmentation 
in RQMD. Further interactions and decays of these hadrons are then 
taken into account in a relativistic transport model.
This model is found to provide a good description of 
hadronic observables in heavy-ion collisions at CERN SPS energies and
does not rely on assumptions of thermal equilibrium
\cite{likob,lib97}.

To calculate the dilepton spectra in heavy-ion collisions, 
we need to know the elementary cross sections for the secondary
processes mentioned above. These processes can be classified into
pseudoscalar-pseudoscalar ($PP$), vector-pseudoscalar ($VP$), 
vector-vector ($VV$), and
axial-vector-pseudoscalar (AP) types. Relying on vector meson 
dominance (VMD), the first three types can be evaluated using effective
hadronic Lagrangians \cite{song94,gale94}.
Concerning the last case, there exist a number of models 
for $\pi\rho a_1$ dynamics.
In Ref. \cite{gao97}, a comparative study was carried out for 
both on-shell properties and dilepton production rates in those  
models. By using the experimentally-constrained spectral
function \cite{huang95} it was found that the effective
chiral Lagrangian of Ref. \cite{comm84} in which the vector mesons
are introduced as massive Yang-Mills fields 
provides satisfactory off-shell, as well as on-shell,
properties for $\pi\rho a_1$ dynamics. The Lagrangian
of Ref. \cite{comm84} will be used in this work in
close conjunction with experimental data, as described below.

The cross sections for these processes have a similar global structure. 
Here we take that for $PP\rightarrow l{\bar l}$ as an
example,
\begin{eqnarray}
\sigma _{PP\rightarrow l{\bar l}} (M)
={8\pi \alpha^2 k \over 3M^3} |F_P(M)|^2 (1-{4m_l^2\over M^2})
(1+{2m_l^2\over M^2}),
\end{eqnarray}
where $k$ is the magnitude of the three-momentum of the pseudoscalar
meson in the center-of-mass frame, $M$ is the mass of the
dilepton pair, $m_l$ the lepton mass, and $|F_P(M)|^2$ 
is the electromagnetic form factor of the process.
The cross sections for $\pi\rho\rightarrow l{\bar l}$ and 
${\bar K}K^*+c.c.\rightarrow l{\bar l}$, which are of the
$VP$ type, have been studied in Ref. \cite{haglin95}.
The cross section for $\pi \omega \rightarrow
l{\bar l} $ has the same form as that for $\pi\rho\rightarrow
l{\bar l}$, but with a different form factor.
Finally the cross section for $\pi a_1\rightarrow l{\bar l}$
is discussed below. 

We emphasize that these cross sections can be constrained
by a wealth of experimental data \cite{eedata} for the reverse process, 
$e^+e^- \rightarrow$ hadrons, via
detailed-balance. One way this is
achieved is by introducing vector meson dominance form factors
which are fitted to the $e^+e^- \rightarrow$ hadrons
data and then used in the meson-meson$\rightarrow l{\bar l}$
processes, as we will do in this work. The pion electromagnetic
form factor is dominated by the $\rho (770)$ meson, 
while that of the kaon is dominated by the $\phi (1020)$ meson. 
At large invariant masses, higher $\rho$-like 
resonances such as $\rho (1450)$ were found to be important.
High isoscalar vector mesons such as $\omega (1420)$ and $\phi (1680)$
play important roles in the electromagnetic form factors
of $\pi\rho\rightarrow l{\bar l}$ and ${\bar K}K^*+c.c.\rightarrow 
l{\bar l}$ \cite{haglin95}. Similarly, the electromagnetic form factor
for $\pi \omega \rightarrow l{\bar l}$ can be measured 
in $e^+e^-\rightarrow \pi^0\pi^0\gamma$.
The extraction of the form
factor for $\pi a_1\rightarrow l{\bar l}$ from $e^+e^-\rightarrow
\pi^+\pi^-\pi^+\pi^-$ does involve some uncertainties concerning 
the $\rho (1700)$ resonance.
As this issue is not totally free of theoretical prejudice
we will consider three scenarios for the 
$\pi a_1\rightarrow l{\bar l}$ form factor: one with and 
one without the $\rho (1700)$ contribution, and one which uses 
a form factor determined from DM2 partial wave analysis data. 
The sensitivity of
the final dimuon spectra to those cases will be assessed.
A clear advantage of the $e^+e^-$ data is that its measurements
cover the invariant mass region we are interested in here. Thus,
no off-shell extrapolations are needed. Note that, however, the
treatment of quantum interference remains a possible
issue in our framework, see, {\it e.g.}, Ref. \cite{peter}.

For dilepton spectra with mass above 1 GeV, the contributions
from charm meson decay and initial Drell-Yan processes
begin to play a role. These hard processes, however,
scale almost linearly with the participant nucleon number,
and can thus be extrapolated from the proton-proton and
proton-nucleus collisions. Such a study has recently been
carried out by Braun-Munzinger {\it et al.} \cite{braun97}.
The results for the central S+W collisions corresponding to
the HELIOS-3 acceptance are shown in the left panel of Fig. 1
and are taken from Ref. \cite{drees96}. These, together
with the dileptons from the decays of primary vector mesons,
are collectively termed `background'. It is seen that these
background sources describe well the dimuon spectra in the
p+W reactions, shown in the figure by solid circles.

However, as can be seen from the figure, the sum of these background
sources grossly underestimates the dimuon yield in central
S+W collisions, shown in the figure by open circles.
Since the dimuon spectra are normalized by the measured
charged particle multiplicity, this underestimation indicates
additional sources of dilepton production in heavy-ion
collisions. Some candidates are QGP and/or
hadronic reactions. So the immediate next step is to check 
whether the contribution from the secondary hadronic processes
can explain this enhancement. For dilepton spectra
at low invariant masses, it is well known that the
$\pi\pi$ annihilation plays an extremely important role in
heavy-ion collisions. It is also expected that the
other secondary processes will play a role in the
dilepton spectra in the intermediate mass region. 

The contributions from the thermal processes outlined above
are shown in the middle panel of Fig. 1. These are 
obtained in the relativistic transport model of Refs. \cite{likob,lib97},
including the HELIOS-3 acceptances, mass resolution, and
normalization \cite{helios}. It is seen that the $\pi a_1$
process is by far the most important source for dimuon
yields in this mass region, as was found in thermal 
rate estimates. The $\pi\omega$ process also play
some role in the entire intermediate mass region, while
the contributions from $\pi\pi$, $\pi\rho$ and $K{\bar K}$ 
are important around 1 GeV. We have also verified that the $AV$
(where $A=a_1$) contributions are suppressed \cite{meis}.

\begin{figure}
\vskip -0.5cm
\begin{center}
\epsfig{file=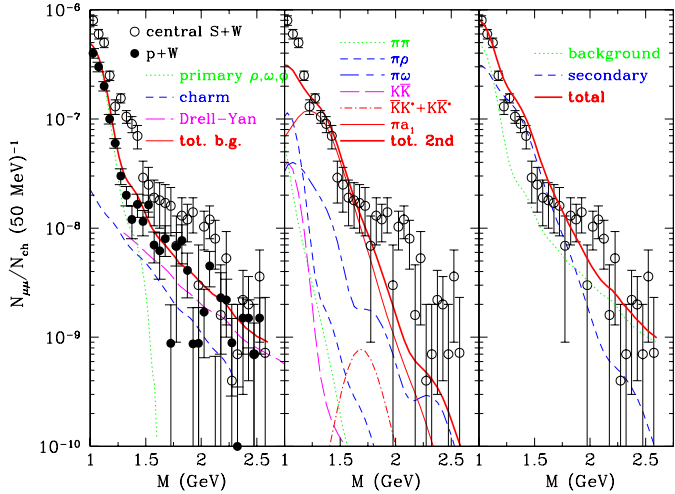,height=4in,width=3in,angle=270}
\caption{Left panel: comparison of background and 
preliminary data in p+W and S+W collisions.
Middle panel: contributions of various secondary
processes of dimuon production in central S+W collisions.
Right panel: comparison of the sum of the background and 
secondary contributions with the preliminary data in
central S+W collisions.}
\end{center}
\end{figure}

In the right panel of Fig. 1, we add the secondary contributions
to the background, and compare
again with the HELIOS-3 data for central S+W collisions. It
is seen that the data can be adequately reproduced. This
highlights for the first time the importance of the secondary processes
for the intermediate-mass dilepton spectra in heavy-ion 
collisions even though the evaluation of the background
itself is not entirely free of complications.
This is an important step forward in the
use of intermediate-mass dilepton spectra as a probe of
the phase transition and QGP formation. Although the current
data do not show any necessity to invoke the QGP formation
in S-induced reactions, consistent with some conclusions 
from $J/\Psi$ physics, we believe that the observation that the secondary
processes do play a significant role in the intermediate-mass
dilepton spectra is interesting and important. We note in passing that
the slight change of the slope observed in the experimental data
corresponds in our interpretation to a crossover 
between the secondary processes and the (Drell-Yan) background. This
signals a passage from hard to soft physics and could lead to
interesting developments. 

In the previous calculation we assumed that all the observed 
4$\pi$ final state in the $e^+e^-$ cross section proceeds through the 
$\pi a_1$ intermediate state. We also did a calculation in which the 
$\pi a_1$ form factor contains only the normal $\rho (770)$. 
The results are shown in the left panel of Fig. 2 by the dotted curve. 
Finally, to complete our survey of possible constraints and 
uncertainties in $\pi a_1 \rightarrow e^+e^-$
cross sections, we did a calculation in which this cross section
is obtained 
from the $e^+e^- \rightarrow \pi a_1$ cross section
determined by the DM2 collaboration in partial wave
analysis (PWA) \cite{dm291}. The results are shown by
dashed curve in Fig. 2. The region between the solid and dashed
curves thus reflect the uncertainty for heavy-ion collisions 
due to our limited knowledge of the $\pi a_1$ cross section. This area is
not unreasonably large.  From a formal point of view, it
is fair to say that no strong evidence currently exists 
coupling the $\rho (1700)$ to a $\pi a_1$ state \cite{pdata}, even 
though better 4$\pi$ data could help resolve this issue along with 
others of interference and $\pi h_1$ contribution \cite{eedata}.

\begin{figure}
\begin{center}
\hskip -1cm
\vskip -0.5cm
\epsfig{file=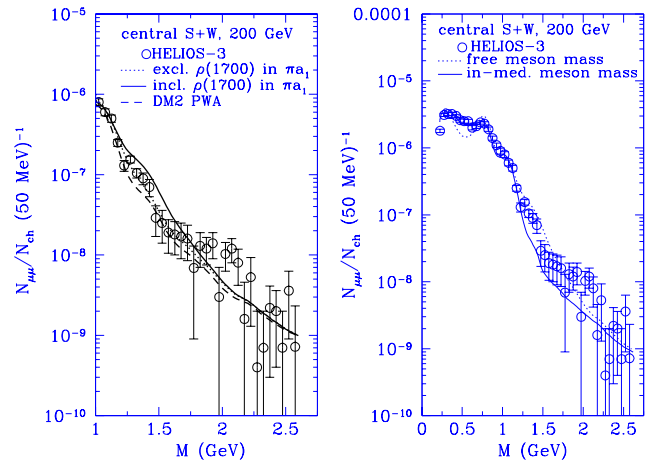,height=4in,width=3in,angle=270}
\caption{Left panel: comparison of dimuon spectra obtained 
with and without the $\rho (1700)$ in the $\pi a_1$
form factor, with those obtained with the DM2 data. Right panel: 
comparison of 
dimuon spectra with
im-medium and vacuum meson masses.}
\end{center}
\end{figure}

Another topic to be adressed here is the effect of dropping vector
meson masses on the entire dimuon spectra from threshold to 
about 2.5 GeV. In Ref. \cite{likob} it was shown that
the enhancement of low-mass dileptons could be interpreted
as a signature of vector meson masses decreasing 
with increasing density and temperature.
This should affect the dilepton spectra in the intermediate-mass
region, mainly through two effects. One is the change of the
invariant energy spectra of these secondary meson pairs. 
The second effect enters through the modification of the 
electromagnetic form factor. Since we can only conjecture how 
the masses of the higher vector resonances change with density 
and temperature, we shall assume for simplicity that they 
experience the same amount of scalar field as the ``common'' rho meson, 
namely, $m_{V,V^\prime} ^* =m_{V,V^\prime}  
-2/3g_\sigma\langle \sigma \rangle$ \cite{likob}. The results of 
this calculation are shown in the right panel of Fig. 2.
Below 1.1 GeV and especially from 0.4 to 0.6 GeV, the agreement 
with the experimental data is much better when the dropping 
vector meson mass scenario is introduced, as was already 
shown in Ref. \cite{likob}. At higher invariant masses the
dropping mass scenarios somewhat underestimates the
experimental data, admittedly not by a large amount. For completeness, 
however, we have to
state that there might be additional contributions from, 
{\it e.g.}, secondary Drell-Yan processes \cite{spie97} that were 
not included in this study. Furthermore, as we progress
higher in invariant mass, the role of baryons has to be 
carefully assessed. So far, the baryons seem to play little
role in the overall dilepton yield \cite{prak97}. This
statement was made \cite{steele97} for masses below 1 GeV,
and is being extended to the intermediate-mass region in \cite{lee97}.
See however Ref. \cite{rapp} for a contrasting viewpoint on the role of
baryons. Finally, the dropping mass curve was obtained using the 
topmost curve of the left panel as a starting point. 

So far, collision broadening is not explicitly included
in our calculation. However, since the ``common'' rho meson is 
treated as a dynamical 
particle in our transport model, its collisions with mesons 
and baryons are included. This shall partially reflect its 
collision broadening 
in hot and dense matter \cite{hag95}. The higher resonances, such as 
$\rho (1450)$ and $\rho (1700)$, are not treated dynamically.
Their effects on dilepton production are included through 
electromagnetic form factors. Therefore, collision broadening 
is not considered explicitly for these higher resonances. However, Since 
they typically have a natural width of 200-300 MeV, a 
moderate broadening width shall not affect our 
results dramatically. This will be dealt with
quantitatively in a more elaborate way in the future.

In summary, we have analysed the recent HELIOS-3 data on 
intermediate-mass dilepton production in heavy-ion collisions 
at CERN SPS energies using a relativistic transport model, 
with the elementary dilepton production cross sections 
constrained by the $e^+e^-$ annihilation data. We have shown 
the importance of secondary processes 
for the dilepton production in heavy-ion collisions in this mass 
region.  

The current investigation can be extended to higher incident energies,
such as those of the RHIC collider, by combining the cross sections
(or thermal rates) obtained in this study with, {\it e.g.}, 
hydrodynamical models for the evolution of heavy-ion collisions
at the RHIC energies. This kind of study is useful for the
determination of hadronic sources in the dilepton 
spectra, and for the clear identification of the dilepton
yield from the QGP.   

\vskip 0.5cm

We are grateful to Nikolai Achasov, Gerry Brown, Sandy Donnachie, 
Madappa Prakash, Ralf Rapp, and 
Ismail Zahed for useful discussions.
This work is supported in part by the U.S. Department of Energy
under grant number DE-FG02-88ER-40388,  by the Natural Sciences
and Engineering Research Council of Canada, and by the Fonds
FCAR of the Qu\'ebec Government.

\end{document}